\documentstyle[prl,aps,multicol,epsf]{revtex}

\begin{document}
\draft
\tighten

\title{Order-out-of-disorder in a gas of elastic quantum strings in 2+1
dimensions.}
\author{J. Zaanen}
\address{Instituut-Lorentz for Theoretical Physics, Leiden University, 
P.O. Box 9506, NL-2300 RA Leiden, The Netherlands}
\date{\today}
\maketitle

\begin{abstract}
A limiting case of a dynamical stripe state which is
 of potential significance to cuprate superconductors is considered:  
a gas of elastic quantum strings in 2+1 dimensions, interacting merely
via a hard-core condition. It is demonstrated that this gas solidifies always,
by a mechanism which is the quantum analogue of the entropic interactions
known from soft condensed matter physics.
\end{abstract}


\begin{multicols}{2}
\narrowtext

The analysis of systems of quantum particles has been traditionally
the focus point of quantum many body theory. On the other hand, much
less is known about systems composed of extended objects. Here I
will analyze one of the simplests examples of such a system: a gas
of quantum strings with finite line tension, embedded in 2+1 dimensional space-time. A motivation to study this problem is found in the context of 
the cuprate stripes\cite{tran}. It is popular to 
view these stripes as preformed line-like textures which can either order 
in a regular pattern, or stay in a disordered state due to strong
quantum fluctuations.   The question arises whether it is possible to
quantum melt a system of completely intact, infinitely long stripes. 
Even in this limit the stripes themselves can still 
execute quantum meandering motions and a consensus has been 
growing that a single stripe is like a quantum string with finite 
line-tension\cite{eskes,morrais,nayak,kivel,hassel}.
I define the ideal string gas as the low density limit
where the width of the strings can be neglected,
while the strings only interact via the requirement that
they cannot intersect\cite{finT}. 
 This is obviously the limit where quantum kinetic energy
is most important. I will show that in 2+1 dimensions even in this limit 
this string system turns into a solid at zero temperature. 
This solidification is driven by the 
quantum-mechanical analogue of the entropic interactions
known from statistical mechanics. In a system with steric interactions
between its constituents, entropy is paid at collisions in the classical
system and kinetic energy in the quantum system.
This causes an effective repulsion and 
these `quantum entropic' interactions dominate to such an extent in
the string gas that they cause it to solidify always.

In the path-integral representation, a quantum-mechanical problem of
interacting particles becomes equivalent to a statistical 
physics  problem of interacting elastic lines (`worldlines'). 
Likewise, the quantum string gas becomes equivalent
to the statistical physics problem of a stack of elastic membranes 
(`worldsheets') which do not interact except for the requirement 
that the membranes do not intersect.  
A seminal contribution in the study of entropic interactions in classical
systems composed of extended entities
is the analysis by Helfrich\cite{helfrich} of a
system of extrinsic curvature membranes in 3D, interacting only via 
an excluded volume constraint. I will illustrate
this method in the quantum context by analyzing 
the hard-core bose gas in 1+1D, which is closely
related to Helfrich's extrinsic curvature membranes in 3D. The string gas 
will turn out to be a straightforward, but non-trivial extension of the bose 
gas: different from the latter, the 
quantum entropic interactions of the string gas are driven by long
wavelength fluctuations.   

To acquire some insights in the Helfrich method in the context of quantum
mechanics, consider the familiar problem of hard-core (but otherwise
non-interacting) bosons in 1+1D. 
This is solved by mapping onto a non-interacting
spinless-fermion gas. Although mathematically trivial, this problem
does exhibit the conceptual ambiguity associated with 
Luttinger liquids\cite{voit}.
On the one hand, it is clearly a gas of particles characterized by a
kinetic scale $E_F$, while at the same time the long wavelength
density-density correlator exhibits the algebraic decay characteristic 
for a harmonic crystal in 1+1D: $< n(x) n(0) > \sim cos (2k_F x) / x^2$.
The concept of entropic interaction offers a simple explanation.   

The hard-core bose gas at zero temperature
corresponds with the statistical physics problem of a gas of non-intersecting
elastic lines embedded in 2D space-time\cite{copper}, which
are directed along the time direction. The space-like 
displacement of the $i$-th worldline is parametrized in terms of a field
$\phi_i (\tau)$ ($\tau$ is imaginary time) and the partition function
is, 
\begin{eqnarray}
Z & = & \Pi_{i=1}^N \Pi_{\tau} \int d \phi_i (\tau) 
e^{-{ S \over {\hbar}} }, \nonumber \\
S & = & \int d\tau \sum_i {{M} \over 2} ( \partial_{\tau}
\phi_i)^2, 
\label{hcZ}
\end{eqnarray}
supplemented by the avoidance condition,
\begin{equation}
\phi_1 < \phi_2 < ...  < \phi_N.
\label{avoid}
\end{equation} 
The hard-core condition Eq.(\ref{avoid}) renders this to be a highly
non-trivial problem. Helfrich considered the related classical problem of
a stack of linearized  and directed extrinsic curvature membranes embedded 
in 3D space. Although this is a higher dimensional problem, the action
depends on double derivatives instead of the single derivatives
in Eq. (\ref{hcZ}), $(\partial_{\mu} \phi)^2 \rightarrow 
(\partial_{\mu}^2 \phi)^2$, and it follows from powercounting 
that this problem is equivalent to the hard-core bose gas in the
present context. In order to determine the `entropic'
elastic modulus at long wavelength Helfrich introduced the following  
construction. Assume that the long wavelength modulus $B_0$ is finite. For
the bose gas this implies
that the long wavelength action is that of a 1+1D harmonic solid,
\begin{equation}
S_{eff}  = { 1 \over 2} \int d\tau \int dx 
\left[ { \rho} (\partial_{\tau} \psi)^2 + B_0 (\partial_{x} \psi)^2 \right],
\label{BSeff}
\end{equation}
where $\psi(x, \tau)$ is a coarse grained long-wavelength displacement field,
$\rho = M / d$ the mass density, and $d$ the average interwordline distance
($n = 1 /d$ is the density). 
Obviously, 
for finite $B_0$ fluctuations are suppressed relative to the case that
$B_0$ vanishes and this cost of kinetic energy in the quantum problem
(entropy in the classical problem) raises the free-energy. Define this
`free-energy of membrane joining' as
\begin{equation}
\Delta F ( B_0) = F (B_0) - F (B_0=0),
\label{delF}
\end{equation}
At the same time, by general principle it has to be that the `true'
long wavelength modulus $B$ in the $x$ direction should satisfy
($V$ is the volume),
\begin{equation}
B = d^2 { { \partial^2 (\Delta F ( B_0) / V)} \over {\partial d^2} }. 
\label{diffeq}
\end{equation}
In case of the steric interactions, the only source of long wavelength 
rigidity in the space direction is the fluctuation contribution to 
$\Delta F$. This means that $B_0 = B$ and $B$ can be self-consistently
determined from the differential equation, Eq. (\ref{diffeq}). In fact,
the only  ambiguity in this procedure is the choice for the short
distance cut-off for the fluctuations in the $x$ direction, which
is expected to be proportional to the distance between the worldlines,
$x_{min} = \eta d$. The shortcoming of the method is that mode-couplings
are completely neglected and this is not quite correct since the outcomes
do depend crucially on short wavelength fluctuations. However, 
it appears\cite{kleinert} 
that these effects can be absorbed in the non-universal `fudge-factor' $\eta$,
giving rise to changes in numerical prefactors
 without affecting the dependence of
$B$ on the dimensionful quantities in the problem.

The free energy difference for the bose gas, Eq. (\ref{delF}),
is easily computed from the Gaussian action Eq. (\ref{BSeff}) and 
expanding up to leading order in  $\lambda = ( \sqrt{B} \tau_0 ) / (
\sqrt{\rho} d)$ ($\tau_0$ is the cut-off time), 
becoming small in the low density limit,
\begin{equation}
{ {\Delta F} \over V}  = 
 { {\pi \hbar} \over {4 \eta^2} } \sqrt{ {B \over M} } { 1 \over {d^{3/2}} }
+ O (\lambda^2).
\label{hacoF}
\end{equation}
Inserting Eq. (\ref{hacoF}) on the r.h.s. of the self-consistency equation 
Eq. (\ref{diffeq}) and solving the differential equation up to leading
order in $\lambda$ yields,
\begin{equation}
B = { {9 \pi^2} \over {\eta^4} } { {\hbar^2} \over {M d^3} }.
\label{Bbosons}
\end{equation}
It is easily checked that this corresponds with the elasticity modulus 
appearing in the bosonized action of the hard-core boson problem, taking
$\eta = \sqrt{6}$. Hence, the space-like rigidity of the bose gas
at long wavelength can be understood as a consequence of entropic 
interactions living in Euclidean space-time.

\begin{figure}[h]
\hspace{0.00 \hsize}
\epsfxsize=0.9\hsize
\epsffile{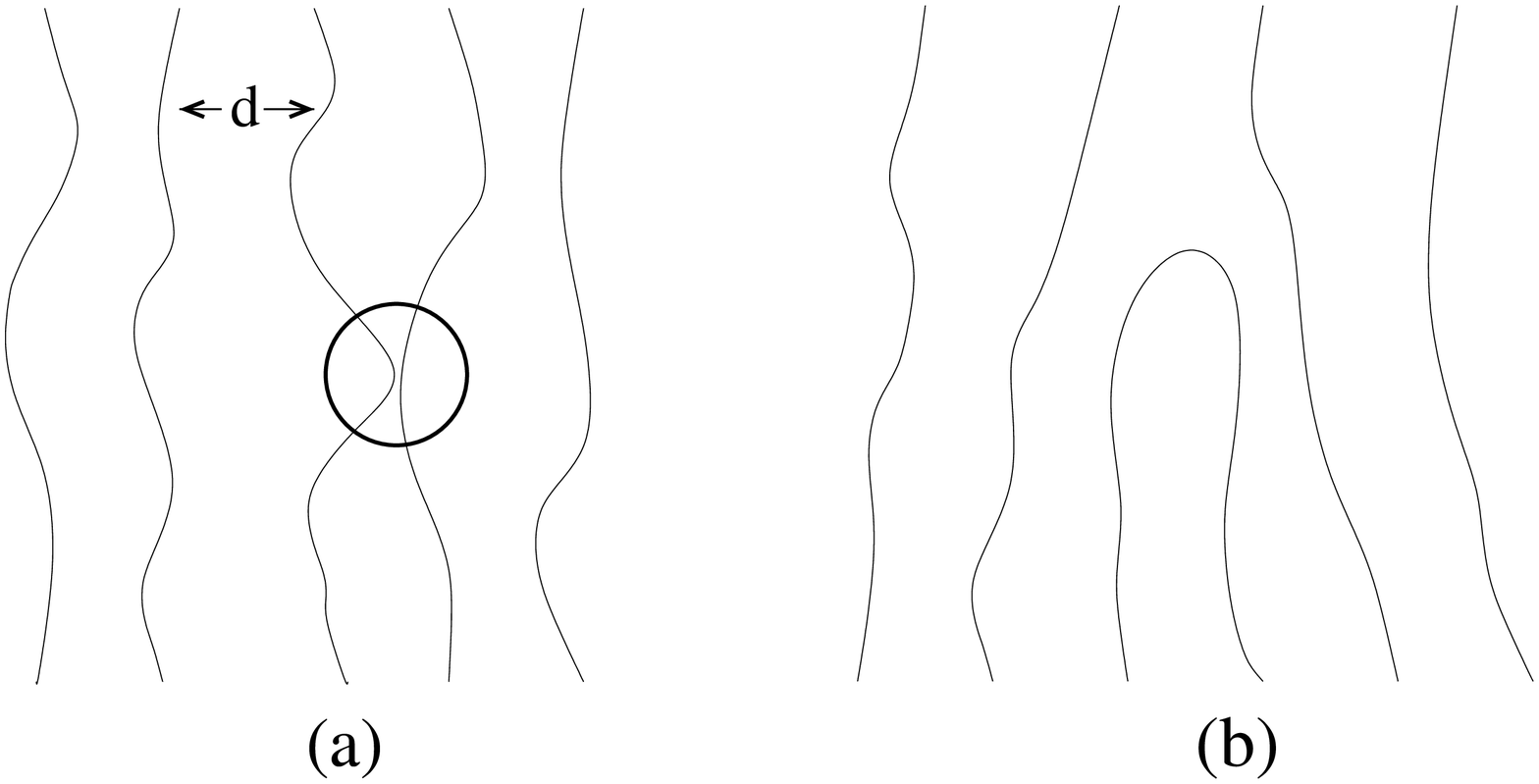}
\caption{ A typical space-like configuration in the directed string gas (a),
including a collision of the type driving the `quantum-entropic'
interactions. In the string gas dislocations (b) do not proliferate,
and it is therefore equivalent to the directed gas.}  
\label{f1}
\end{figure}

Let us now turn to the string-gas problem. In fact, the string gas in 2+1D is
related to the hard-core bose gas in 1+1D: the latter can be viewed
as the {\em compactified} version of the former. Imagine that the hard-core
bose gas lives actually in 2+1D where the additional dimension $y$ is
rolled up to a cylinder with a compactification radius $R_y$ of order of
the lattice constant $a$, while
the bosons are spread out in elastic strings wrapped 
around the y-axis. Let $R_y$ go to infinite. This has the effect that
the embedding space becomes $2+1$ dimensional, while the boson wordlines
spread out in string worldsheets. This `directed string-gas' is
not yet the one of interest, since the worldsheets
are not only directed along the imaginary time directions (as required by
quantum mechanics) but also in the $x-y$ plane (Fig. 1a). The
difficulty is that in the string gas dislocations 
can occur (Fig. 1b), and if these proliferate
they will destroy the generic long range order of the directed string gas.
However, two objections can be raised against a dislocation
mediated quantum melting. The first objection involves a further specification:
already a single string tends to acquire spontaneously a direction, if
it is regularized on a {\em lattice} (like the stripes).  As 
pointed out by Eskes {\em et. al.}\cite{eskes}, 
the reason is that `overhangs' like
in Fig. 1b are events where transversal fluctuations are surpressed,
relative to those around directed configurations. 
The second argument is more general.  It is
a classic result\cite{toprov,finT} 
that at any finite temperature dislocations proliferate
in the string gas. However, in the presence of a
finite range interaction of any strength  the Kosterlitz-Thouless transition
will occur at a finite temperature. Hence, by letting this interaction to
become arbitrary weak a $T=0$ transition can be always circumvented.
    
When dislocations can be excluded the directed
string gas remains and this is just the decompactified Bose gas.   
In Euclidean space-time it corresponds with a sequentially ordered stack 
of elastic membranes. Orienting the worldsheets in the $y, \tau$ planes,
the action becomes in terms of the dispacement fields $\phi_i(y,\tau)$
describing the motion of the strings in the $x$ direction,  
\begin{eqnarray}
Z & = & \Pi_{i=1}^N \Pi_{y,\tau} \int d \phi_i (y, \tau) 
e^{-{ {\vec{S}} \over {\hbar}} }, \nonumber \\
S & = & \int d\tau dy \sum_i \left[ {{\rho_c} \over 2} ( \partial_{\tau}
\phi_i)^2 +  {{\Sigma_c} \over 2} ( \partial_y \phi_i )^2 \right], 
\label{strZ}
\end{eqnarray}
again supplemented by the avoidance condition Eq. (\ref{avoid}).
In Eq. (\ref{strZ}), $\rho_c$ is the mass density and $\Sigma_c$ the string 
tension, such that $c = \sqrt{ \Sigma_c / \rho_c}$ is the velocity.
In the remainder
I choose a lattice regularization with lattice constant $a$, such that 
the average string-string 
distance is $d = a / n$ ($n$ is the density), and UV momenta and frequency cut-offs on
a single string
$q_0 = \pi / a$ and $\omega_0 = c q_0$, respectively.

Turning to the Helfrich method, the effective long wave-length action
is written as,
\begin{equation}
S_{eff}  = { 1 \over 2} \int d\tau dx dy
\left[ \rho (\partial_{\tau} \psi)^2 + B (\partial_{x} \psi)^2 + 
 \Sigma(\partial_{y} \psi)^2 \right],
\label{Seff}
\end{equation}
with $\psi(x,y,\tau)$ as the coarse grained long-wavelength displacement field,
while $\Sigma = \Sigma_c / d$, $\rho = \rho_c / d$
and $B$ has to be determined.     
From the action Eq. (\ref{Seff}) it follows that the free energy
difference Eq. (\ref{delF}) is,
\begin{equation} 
{ {\Delta F} \over V}  =  - { {\hbar c} \over { 8 \pi^2} } \int_0^{q_0} dq^2
\int_0^{\pi/(\eta d)} dq_x \ln \left[ {  {\Sigma q^2} \over
{\Sigma q^2 + B q^2_x} } \right].
\label{delFst}
\end{equation}
This integral is easily solved analytically and expanding in
the small parameter $\lambda = ( \sqrt{  B }  a )/
(\sqrt{\Sigma} \eta d )$,
\begin{equation}
{ {\Delta F} \over V} = { { \pi \hbar c} \over {24 \eta^3 \Sigma_c} }
( { B \over {d^2} }) ( { 5 \over 3} +  \ln \left[ 
{ { \eta^2 \Sigma_c} \over {a^2} } { d \over B } \right] ) + 
O (\lambda^4).
\label{delFsa}
\end{equation}
The free-energy difference is proportional to $B/ d^2$ except for a 
logarithmic `correction' $\sim ( B / d^2) \ln (d / B)$. Since $B$ tends to 
zero in the low density limit, it is actually this logarithmic `correction'
which determines the low density asymptote of the differential 
equation which is obtained after substition of Eq.(\ref{delFsa}) in the
self-consistency condition Eq. (\ref{diffeq}).  The physical meaning
of this logarithm will be discussed later.

The differential equation determining the fluctuation induced
modulus $B$ becomes,
\begin{equation}
f (d) = - C_0 { {\partial^2} \over {\partial d^2} }
\left[ f (d) \ln ( C_1 d f ( d ) ) + { 5 \over 3} \right],
\label{diffe}
\end{equation}
where $ f ( d ) = B / d^2$ and 
$ C_0 = ( \pi \hbar c) / ( 24 \eta^3 \Sigma_c), C_1 = a^2 / (\eta^2
\Sigma_c)$. Eq.(\ref{diffe}) can be simplified using 
the Ansatz $ f ( d ) = exp [ - \Phi (d) ]$. It is easy to see that
for large $d$ the second derivative terms  $\sim \partial_d^2 \Phi$ 
can be neglected relatively to the first derivative terms
(`quasiclassical approximation'). Neglecting the other terms which 
do not contribute in the low density asymptote (including the one
derived from the `$5/3$' term in Eq. (\ref{diffe})) $\Phi$ obeys
asymptotically the simple differential equation,
\begin{equation}
( \Phi - 2 ) ( { {\partial \Phi} \over {\partial d} } )^2
= { 1 \over {C_0} },
\label{phidif}
\end{equation}
and it follows that $\Phi ( d ) \sim d^{2/3}$. The full expression
for the induced modulus is up to leading order in the density,
\begin{equation}
B = A d^2 e^{ - \eta ({ 54 \over {\pi}})^{1/3} { 1 \over {\mu^{1/3} } } },
\label{Bfinal}
\end{equation}
where $A$ is an integration constant and $\mu$ is the
`coupling constant' for the string-gas,
\begin{equation}
\mu = { {\hbar} \over { \rho c d^2} }.
\label{mu}
\end{equation}
Eq.'s (\ref{Bfinal},\ref{mu}) represent my central result.

What is the significance of this result? Most importantly, it
demonstrates that in parallel with the hard-core bose gas (and
Helfrich's membranes), the string 
gas is characterized by a fluctuation induced elastic modulus
at long wavelength which will be small but finite even at low
density. This modulus $B$ appears in the action Eq. (\ref{Seff})
which describes an elastic manifold covering 2+1D space-time.
Eq. (\ref{Bfinal}) describes the counter-intuitive fact that
upon increasing the kinetic energy of a single string, the
rigidity of this medium  is actually increasing. 
The parameter $\mu$ is the dimensionless quantity measuring the importance
of quantum fluctuations\cite{zahovs}.  
In order to prohibit diverging fluctuations
on the lattice scale, $\mu$ should be less than one, while the
classical limit is approached when $\mu \rightarrow 0$. According to
Eq. (\ref{Bfinal}), $B$ depends on $\mu$ in a 
stretched exponential form, such that $B$ increases when $\mu$
is increasing. Since quantum dislocation
melting is prohibited,  the string gas is always
a solid, and this solid becomes more rigid 
when the microscopic quantum fluctuations become more important. This might
appear as less surprising when the (directed) string gas is viewed
as a decompactified bose gas. On the one hand, the larger internal
dimensionality of the worldsheets as compared to the worldlines 
weakens the `quantum-entropic' interactions, but the enlarged
overall dimensionality causes the algebraic long range order
of the 1+1D bose gas to become the true long range order of the  
2+1D string gas.

The mechanism behind the `quantum entropic' interaction is  actually
different from the one in the bose gas. In the bose gas
 it builds up at short-wavelengths, while
in the string gas it is
driven by the long-wavelength fluctuations living on the strings. 
An alternative, more
intuitive understanding is available for the Bose-gas result,
Eq. (\ref{Bbosons}). 
This is based on the simple notion that
every time membranes/worldlines collide an amount of 
entropy $\sim k_B$ is paid because the membranes 
cannot intersect\cite{helfrich,fischer,copper}.
Hence, these collisions raise the free energy of the system and
this characteristic free energy cost $\Delta F_{coll} \sim
k_b T n_{coll.}$. The density of collisions $n_{col}$ is easily
calculated: for the worldlines, the mean-square transversal
displacement as function of (time-like) arclength increases
like $\langle (\phi(\tau) - \phi (0) )^2 \rangle = ( \hbar / M) \tau$.
The characteristic time $\tau_c$ it takes for one collision to occur is
obtained by imposing that this quantity becomes of order $d^2$ and 
a characteristic collision energy scale is obtained $E_F \sim 
\hbar / \tau_c \sim (\hbar^2/ M) n^2$.
$E_F$ is of course the Fermi-energy: it is the scale separating a
regime where worldlines are effectively isolated ($E > E_F$, free
particles) form one dominated by the collisions ($E < E_F$, Luttinger
liquid). The induced modulus follows
from naive coarse graining: $E_F$ is the characteristic energy asssociated with
density change, while $d$ is the characteristic length. Therefore,
 $B \sim E_F/d$, reproducing the result
Eq. (\ref{Bbosons}), within a prefactor of order unity.

For the string gas this procedure yields a simple exponential-
instead of the stretched exponential Eq. (\ref{Bfinal}). 
The mean-square transversal displacement now depends 
logarithmically of the worldsheet area $A$: 
$\langle (\Delta \phi (A))^2 \rangle = \hbar /(\rho c) \ln (A)$.
Demanding this to be equal to $d^2$, the degeneracy scale 
follows immediately. The characteristic worldsheet area $A_c$
for which on average one collision occurs is given by 
$\hbar/(\rho c) \ln (A_c) \simeq d^2$
where $A_c = c^2 \tau^2_c / a^2$ in terms of the collision time $\tau_c$.
It follows that $\tau_c \simeq (a/c) e^{1 /2\mu}$ and the `Fermi energy'
of the string gas is of order
 $E_F^{str} = \hbar / \tau_c \simeq (\hbar c / a)
\exp { (-1 / 2\mu)}$ and thereby $B \sim \exp{ (- 1 / \mu)}$.
In fact, the same $\mu$ dependence is obtained from 
Helfrich's method if the logarithm in Eq. (\ref{delFsa}) is neglected. 
Hence, this collision picture misses entirely the origin of the
quantum-entropic repulsions in the string gas!
The reason becomes clear by inspecting the origin of the
logarithmic term in the integrations leading to Eq. (\ref{delFsa}).
Cutting off the smallest allowed momenta in the $x, \tau$ directions
by $q_{min}$ one finds that $\ln [ (\eta^2 \Sigma_c d) / ( a^2 B) ] 
\rightarrow - \ln [ (a^2 B) / (\eta^2 \Sigma_c d) + a^2 q^2_{min} ]$, 
and this is unimportant for any finite 
$q_{\min}$ in the low density limit. Therefore, the logarithm
and thereby the induced modulus are driven by the long wavelength
fluctuations on the strings, and these are not considered in the
collision point picture.

In summary, I have analyzed the fluctuation induced interactions
in the `ideal' gas of elastic quantum strings in 2+1D. A novelty
is that in this system the induced elasticity is due to long wavelength 
fluctuations, qualitatively different from  the short distance
physics of the Bose gas. It remains to be seen if these interactions are 
of relevance in real physical systems. On the one hand, these are rather 
weak and  easily overwhelmed by direct interactions\cite{tsvelik}. 
However, 
direct string-string interactions which decay exponentially are generic and
in this case the induced interactions can dominate at sufficiently low
density because of their stretched exponential dependence on density:
in principle the induced interactions can be physical observables. 
The immediate relevance of my findings lies elsewhere. The strings considered
here are idealizations of the stripes but these idealizations are nevertheless 
close to a popular way of viewing these matters. I have demonstrated
that in the absence of zero temperature stripe 
long range order\cite{smectic} it has to be
that these ideal stripes are broken up in one or the other way.     
       
Many stimulating interactions are acknowledged with S. I. Mukhin and  
W. van Saarloos.
I thank P. van Baal for his suggestion to study the literature on 
extrinsic curvature membranes.

\end{multicols}


\begin{references}
\bibitem{tran}
J.M. Tranquada {\em et. al.}, Nature {\bf 375}, 561 (1995). 
\bibitem{eskes}
H. Eskes {\em et al}, Phys. Rev. B {\bf 58}, 6963 (1998);
J. Zaanen, O. Y. Osman, and W. van Saarloos, Phys. Rev. B {\bf 58}, R11868
(1998).
\bibitem{morrais}
C. Morais Smith {\em et. al.}, Phys. Rev. B {\bf 58}, 453 (1998). 
\bibitem{nayak}
Y. Nayak and F. Wilczek, Phys. Rev. Lett. {\bf 78}, 2465 (1997).
\bibitem{kivel}
S. A. Kivelson, E. Fradkin and V. J. Emery, Nature {\bf 393}, 550 (1998). 
\bibitem{hassel}
N. Hasselmann {\em et. al.}, Phys. Rev. Lett. {\bf 82}, 2135 (1999);
Y. A. Dimashko {\em et. al.}, Phys. Rev. B {\bf 60}, 88 (1999).
\bibitem{finT}
The classical, finite temperature version of this problem has been
analyzed thoroughly in the past (see \cite{copper} and ref.'s therein) and
considered in the context of stripes in \cite{zahovs}.
\bibitem{copper}
S. N. Coppersmith {\em et. al.}, Phys. Rev. B {\bf 25}, 349 (1982). 
\bibitem{zahovs} J. Zaanen, M. Horbach and W. van Saarloos,
Phys. Rev. B {\bf 53}, 8671 (1996).
\bibitem{helfrich}
W. Helfrich, Z. Naturforsch. A 33, 305 (1978); W. Helfrich and R. M.
Servuss, Nuovo Cimento D {\bf 3}, 137 (1984).
\bibitem{voit}
J. Voit, Rep. Prog. Phys. {\bf 57}, 977 (1994), and references therein.
\bibitem{kleinert}
W. Janke and H. Kleinert, Phys. Rev. Lett. {\bf 58}, 144 (1987);
W. Janke, H. Kleinert and M. Meinhardt, Phys. Lett. B {\bf 217}, 525 (1989);
see also H. Kleinert, Phys. Lett. A {\bf 257}, 269 (1999).
\bibitem{toprov}
V. L. Pokrovsky and A. L. Talapov, Phys. Rev. Lett. {\bf 42}, 65 (1979).
\bibitem{fischer}
M. E. Fisher and  D. S. Fisher, Phys. Rev. B {\bf 25}, 3192 (1982).
\bibitem{tsvelik}
For instance, the Toda Array corresponding with a soft-core version of
the string gas with exponential repulsions does not show quantum
entropic interactions: C. P\'epin and A. M. Tsvelik, 
Phys. Rev. Lett. {\bf 82}, 3859 (1999).
\bibitem{smectic}
Notice that the quantum smectic as proposed by Kivelson 
{\em et. al.}\cite{kivel} is a solid in the present context. 

\end{references}
\end{document}